\documentclass[aps,prl,superscriptaddress,showpacs,floatfix,reprint]{revtex4-1} 

\usepackage{amsfonts}
\usepackage[caption=false]{subfig}
\usepackage{amssymb}
\usepackage{amsmath}
\usepackage{commath}
\usepackage{graphicx,bm}
\usepackage{verbatim}
\usepackage{color}
\usepackage{multirow}
\renewcommand{\v}[1]{\ensuremath{\mathbf{#1}}} 
 
\let\baraccent=\= 
\renewcommand{\=}[1]{\stackrel{#1}{=}} 

\begin{document}

\title{Berezinskii-Kosterlitz-Thouless transition in the XY model on the honeycomb lattice: A comprehensive Monte Carlo analysis}

\author{Felipe E. F. de Andrade}
\affiliation{Instituto Federal de Goi{\'a}s, Rua 76, Centro, Goi{\^a}nia - GO, Brazil}
\author{L. N. Jorge}
\affiliation{Instituto Federal do Mato Grosso - Campus C{\'a}ceres,
Av. dos Ramires s/n, 78201-380, C{\'a}ceres, MT, Brazil}
\author{Claudio J. DaSilva}
\email{claudio.silva@ifg.edu.br}
\affiliation{Instituto Federal de Goi{\'a}s, Rua 76, Centro, Goi{\^a}nia - GO, Brazil}

\date{\today}

\begin{abstract}
In this paper, we thoroughly examined the Berezinskii-Kosterlitz-Thouless (BKT) phase transition in the two-dimensional XY model on the honeycomb lattice. To address its thermodynamical behavior, we combined standard numerical Monte Carlo simulations with the simulated annealing (SA) protocol and entropic simulations based on the Wang-Landau (WL) algorithm. The transition temperature was determined using the second ($\Upsilon$) and fourth-order ($\Upsilon_4$) helicity modulus as the order parameter. Our best finite-size scaling analysis suggests $T_{BKT} = 0.575(8)$ from SA and $T_{BKT}=0.576(3)$ from WL. These values deviate significantly from the expected theoretical value of $1/\sqrt{2}$. We believe that this discrepancy suggests that the theoretical assumptions regarding the analytical calculation may need to be revisited. Additionally, we calculated the vortex density and the formation energy of the vortex-antivortex pairs, where the obtained vortex formation energy is $2\mu=5.80(12)$. Upon comparison with the square lattice, our results support the notion of instability of the honeycomb lattice to support the spin long-range order and give additional backing to the critical behavior we found.
\end{abstract}

\maketitle 

\section{Introduction}

In 2016, the Nobel Prize in Physics was awarded to David J. Thouless, F. Duncan M. Haldane, and J. Michael Kosterlitz for their groundbreaking contributions to the understanding of topological phase transitions and phases of matter, which have significantly advanced condensed matter physics. They demonstrated that, at sufficiently low temperatures, the magnetic moments of a two-dimensional lattice of atoms form vortex-antivortex pairs due to the presence of topological defects. However, a topological phase transition occurs when the system temperature increases. At a certain critical temperature ($T_{BKT}$), these pairs dissociate and roam freely through the lattice. This transition is called Kosterlitz-Thouless (KT) transition or BKT, in honor of Vadim Berezinskii, a deceased theoretical physicist who presented ideas similar to Kosterlitz and Thouless\cite{Berezinski1972}. 

The two-dimensional XY model with continuous symmetry is the most commonly used theoretical model to study the BKT transition. It exhibits a quasi-long-range ordered phase with the transition not resulting in spontaneous symmetry breaking as expected for ferromagnetic or antiferromagnetic models in two dimensions\cite{Kosterlitz1973,Kosterlitz_1974}. The model has broad applicability, as it can describe various physical systems, such as multicomponent superconductors\cite{BABAEV2004,Hlubina2008} and nematic or smectic liquid crystals\cite{Bruinsma1982,Aeppli1984}. Several analytical and numerical analyzes have been conducted for decades to study the main characteristics of this model, with a significant focus on providing a precise description of the nature of the BKT transition\cite{Minnhagen2003}. Numerous variations and generalizations of this model have also been explored\cite{Deng_2014,Lach2021,Sun2022}, such as incorporating competition between ferromagnetic and nematic interactions and introducing different types of spin interactions\cite{LandauDMXY2022,Cui2022}. These variations have paved the way for the investigation of new thermodynamic phases and topological defects. Despite significant advancements, the model continues to present challenges that require further research\cite{Touchette2022}.

The present study is motivated by two key factors. First, the BKT transition temperature for the square lattice is well established in the literature, with numerical calculations yielding $0.8935(1)$\cite{Hsieh_2013}. For the honeycomb lattice, some analytical studies predict an exact value of $1/\sqrt{2}\approx0.707$\cite{Nienhuis1982,campostrini1996}, which is confirmed by later calculations using cluster simulations\cite{Deng2007} and diffusion map method\cite{jielin2021}. Nonetheless, a detailed Monte Carlo (MC) study of the critical behavior on the honeycomb lattice appears to be absent in the existing literature. The second motivation stems from the ongoing debate regarding the appropriate parameter to characterize the BKT transition, given the limitations of computer simulations due to the finite size of simulated systems. The modulus of second-order helicity, which measures the system response to a global twist of spins in a given direction, is currently accepted as the parameter to describe this transition\cite{Minnhagen2003}. The controversy arises because the discontinuity in the helicity modulus is not strongly dependent on the system size. While some researchers have shown that logarithmic corrections applied to the helicity modulus can yield accurate estimates of $T_{BKT}$\cite{Hsieh_2013}, it is also possible to use fourth-order helicity as an order parameter. This alternative presents a discontinuity at the critical temperature, potentially providing a more precise description of the BKT transition via entropic numerical simulations, where a high resolution search for the critical points is possible\cite{Caparica2014a}.

We primarily use the MC method and the Simulated Annealing (SA) protocol\cite{landau_binder_2014} to calculate the second- and fourth-order helicity. This helps us to estimate the inverse critical temperature $T_{BKT}$. Additionally, we use entropic simulations based on the Wang-Landau (WL) algorithm\cite{Wang2001a} to validate the SA results. The WL technique allows for a more precise determination of the critical region without being affected by critical slowdown. However, it is limited to small system sizes for models with continuous symmetry. Moreover, we will employ the finite-size scaling technique with logarithmic correction proposed by Y.D. Hsieh et al.\cite{Hsieh_2013}. We also calculate the vortex density and the energy formation of vortex pairs, which may provide insight into the stability of the honeycomb lattice compared to the square lattice and the location of the critical region.

This paper is organized as follows. First, in section II we review the thermodynamics of the XY model and the BKT transition. In Section III we then detail the numerical procedure used and the calculation of the thermodynamic properties. Next, in section IV we discuss the results. Finally, in section V we draw some conclusions.

\section{The model}

The XY model represents a system of spins arranged at the sites of a two-dimensional lattice\cite{Kosterlitz1973} and serves as a powerful tool for modeling magnetic systems in two dimensions, such as magnetic crystals\cite{Gong2019} and van der Waals heterostructures\cite{Gong2017}.

The Hamiltonian of the standard XY model can be written as
\begin{align}
{\cal H} = -J \sum_{\langle i,j \rangle}\v{S}_i\cdot\v{S_j}= -J \sum_{\langle i,j \rangle} \cos(\theta_i -\theta_j),
\label{eq:hamil}
\end{align}
where $J>0$ is the ferromagnetic coupling (exchange interaction) constant, $\v{S}_i=(S_i^x,S_i^y)=(\cos\theta_i,\sin\theta_i)$ represents the spin vector $i$, $\theta_i -\theta_j$ is the angle between each pair of spins located in the lattice sites and the notation $\langle i,j \rangle$ indicates that the sum is taken only over nearest-neighbors. The scientific community has been studying phase transitions in models like this for many years. This is largely due to the presence of topological order\cite{Kosterlitz1973}. 

Usually, phase transitions are characterized by order parameters, which are zero in one phase and finite in the other, such as magnetization in ferromagnetic materials\cite{landau_binder_2014}. Close to the critical temperature $T_c$, the thermodynamic properties present a critical behavior that can be described by power laws governed by critical exponents\cite{stanley1971phase}. 

However, in the XY model, there is no discontinuity in the first derivatives of the free energy at the transition temperature, but rather in the second derivatives. Furthermore, there is no spontaneous symmetry breaking of the system, but rather a change in the nature of quasi-long-range fluctuations of the order parameter, i.e., there are no abrupt changes of the system during the transition. This configures a phase transition of continuous (infinite) order\cite{Berezinski1972, Kosterlitz_1974, Kosterlitz1973}. For sufficiently low temperatures, there is the creation of vortex-antivortex pairs in the lattice, representing a state with almost long-range order, as predicted by Kosterlitz and Thouless\cite{Kosterlitz1973}. Vortices can be characterized as topological defects, where the spins arrange themselves in a whirlpool-like formation, that is, when adding the phase differences between the spins in a closed path proportional to the lattice geometry, it yields a value multiple of $\pm 2\pi$. As the temperature increases, a transition from this state to a disordered one occurs, breaking the vortex-antivortex pairs, which begin to move freely through the system. 
 
\section{Numerical procedure}

\subsection{The Metropolis Algorithm}

The MC method uses random numbers to construct the phase space of a given physical system, which can be studied by sampling its microstates. This capability of the MC method is convenient when one wants to represent complex and random systems with fair precision\cite{landau_binder_2014}.

In this sense, to calculate the average value of a quantity of interest in a system in thermal equilibrium, we could use the MC method to generate several states to measure that quantity in each of them. However, we know that the probability of a system in thermal equilibrium occupying a state $i$ with energy $E_i$ can be obtained through the Boltzmann distribution
\begin{align}
  {\cal P}(E_i) = \frac{\exp(-\beta E_i)}{Z},
\label{eq:boltz}
\end{align}
where $\beta = 1/k_BT$, with $k_B$ being the Boltzmann constant, $T$ the temperature, and $Z$ the partition function.

However, generating completely random and independent states is not a coherent representation of natural physical processes. Therefore, we use the Markov chain method to generate new states. In this method, a state $i$ can only evolve into a state $j$ if a transition probability is satisfied, that is, a new configuration cannot be chosen randomly among all possible ones, but rather from the configuration in which the system is located.

Thus, we can represent this transition probability as the ratio between the probabilities of the states $i$ and $j$, that is,
\begin{align}
  \frac{{\cal P}(E_j)}{{\cal P}(E_i)} = \frac{\exp(-\beta E_j)/Z}{\exp(-\beta E_i)/Z} = \exp(-\beta \Delta E),
\label{eq:markov}
\end{align}
where $\Delta E = E_j - E_i$ is the energy difference between these states.

In one of the numerical procedures used in this paper, each new configuration consists of assigning a new orientation to a single spin of the lattice, if this new orientation contributes to the reduction of the system's energy or does not change it, that is, $\Delta E \leq 0$ , the new state is accepted. However, if there is an increase in the energy of the system ($\Delta E > 0$), there is a chance that the new generated state will be accepted according to Equation \ref{eq:markov}. This protocol is known as the Metropolis algorithm, and can be implemented in MC simulations to generate configurations through a probability of acceptance of the new spin orientation ($P_a$) as follows\cite{NewmanCP2012}
\begin{align}
P_a =
\begin{cases}
1 & \text{se} ~\Delta E \leq 0,\\
\exp(-\beta \Delta E) & \text{se} ~\Delta E > 0.
\end{cases}
\end{align}
By scanning the entire lattice, that is, all $N$ sites, where $N = L^2$, assigning a new orientation to the spin at each site and using the Metropolis algorithm to accept or not the new orientation, we perform what is known as the MC sweep ($m_s$).

\subsection{Simulated Annealing}

Due to the complexity of the phase space of the XY model, the Metropolis algorithm might face challenges in sampling the low-temperature region. An effective technique to address this issue is the \textit{Simulated Annealing}\cite{landau_binder_2014}. This technique consists of initializing the system at an infinite temperature and cooling it uniformly and slowly until $T \approx 0$. This protocol is efficient as it enables the system not to get stuck in local energy minima when minimizing a function dependent on numerous parameters. Therefore, the SA procedure allows the system to pass through local minima without getting stuck in them, as long as the increase in $\beta=1/T$ is small in each interval.

In practice, we start with $\beta = 0$ and increase it uniformly until we reach a desired temperature value. Each simulation consists of $n$ temperature intervals, where within each interval $m_s$ MC sweeps are performed, resulting in a total of $nm_s$ simulation steps. After each MC step, the increments in $\beta$ are given by $\beta=1/k_BT=\beta_0+i\Delta\beta_n \quad (i=1,...,nm_s)$, where $ \Delta\beta_n=(\beta_n-\beta_0)/nm_s$.

\subsection{Wang-Landau simulation}

In the study of phase transitions in physical systems, entropic simulations have proven to be an extremely useful tool\cite{Caparica2015,Caparica2012}. Based on the Wang-Landau scheme, they are performed by a random walk in energy space, driven by an unbiased change in spin states, to construct the density of states $g(E)$. The probability of accepting a new spin state is proportional to the reciprocal of the density of states, $1/g(E)$. During the simulation run, a histogram of energy values, $H(E)$, is accumulated, and the usual flatness criterion for $H(E)$ is set $90\%$. After estimating the density of states, these simulations allow the calculation of average values of relevant thermodynamic quantities at any temperature, significantly expanding the information available about the system. In our work, we use entropic sampling based on the WL method\cite{Wang2001a,Wang2001}, with some improvements, such as the adoption of the MC sweep to update the density of states, the accumulation of microcanonical average is done only after a more advanced WL level, the use of an automatic criterion to end the simulation run, and the use of an improvement that saves computational time, which consists of starting all the rounds from a single previous simulation up to some WL level and performing all the other runs from there\cite{Caparica2012,Caparica2014a,Caparica2015}.

A single simulation is performed up to $f_6$ for all lattice sizes, followed by multiple independent simulations from that point onward. This procedure is justified because the density of states remains uncorrelated up to $f_6$, allowing independent results to be obtained, comparable to those generated from $f_0$ \cite{ferreira2018}. This approach reduces computational time by approximately $60\%$ and also ensures that microcanonical averages are accumulated starting from the eighth WL level.

Once the density of states $g(E)$ is obtained, the canonical average of any thermodynamic quantity $X$ can be calculated as:
\begin{equation}
\langle X \rangle_T = \frac{\sum_E \langle X \rangle_E g(e) e^{-\beta E} }{\sum_E g(e) e^{-\beta E}},
\end{equation}
where $\langle X \rangle _E$ is the microcanonical average accumulated during the simulations.

\subsection{Thermodynamic quantities}

To characterize the thermodynamic behavior of the XY model, we calculated the average energy per site, magnetization, and the associated thermal fluctuations, including specific heat and susceptibility. We also analyzed the behavior of the fourth-order cumulant of the order parameter. The average energy in this case is given by:
\begin{align}
\langle E \rangle = \frac{1}{N} \langle {\cal H} \rangle _T,
\label{eq:ener}
\end{align}
where $N$ is the total number of spins, $\cal H$ is the Hamiltonian of the system given by Eq. \ref{eq:hamil}, and its average is taken at a fixed temperature.

Although magnetization is not the most suitable order parameter for determining the BKT transition temperature, it plays a significant role in qualitatively understanding the phase transition. Therefore, the magnetization of the XY model on a finite lattice can be expressed as:
\begin{align}
m=\langle |\v{M}|\rangle_T =\frac{1}{N} \langle \sqrt{M_x^2+M_y^2}\rangle_T,
\label{eq:mag}
\end{align}
where
\begin{align}
M_x =\sum_{i=1}^N\cos\theta_i, \\
M_y =\sum_{i=1}^N\sin\theta_i.
\end{align}
From Eq. \ref{eq:mag} it can be inferred that $m=1$ corresponds to the ground state and is expected to decrease as the temperature rises. For the specific heat and susceptibility we have,
\begin{align}
     C_v = \frac{1}{k_BT^2}(\langle E^2\rangle - \langle E\rangle ^2).
\end{align}
\begin{align}
     \chi = \frac{1}{k_BT}(\langle M^2\rangle - \langle M\rangle ^2).
\end{align}

It is suggested that the second-order helicity modulus serves as an appropriate order parameter to investigate the BKT transition\cite{Minnhagen2003, Hsieh_2013}. This quantity represents the response of the system to a global twist applied to the spins along a specific direction. To better conceptualize helicity, consider a square lattice of spins at a temperature $T = 0$ in a ferromagnetic phase, with linear size $L$, and subjected to a magnetic field parallel to one of its edges. The spins located at the edge that is under the action of the magnetic field must undergo a change of $\Phi$ in their orientation relative to the original state. To minimize the system energy, each spin of neighboring rows parallel to the edge will undergo a rotation equivalent to an angle of $\phi = \Phi / (L - 1)$. Thus, the energy in the presence of this spin is given by $E(\phi) = -J ( L - 1 ) L \cos(\phi)= E(0) + J ( L-1 )L [1 - \cos(\phi) ]$. According to \cite{Sandvik_2010}, for a small $\phi$, we have $E(\phi) - E(0) = (J/2)(L-1)L\phi^2$. Using this result, one can see that the change in the energy for a small twist corresponds to a function of $\phi^2$, thus, the helicity for $T=0$ is defined as
\begin{align}
\langle \Upsilon \rangle_{T = 0} \equiv \frac{\partial^2E(\phi)}{\partial \phi^2}.
\end{align}

For a non-zero temperature, the system is described by the free energy $F(\phi)$, whose global minimum corresponds to $\phi = 0$. Thus, $F(\phi)$ around $\phi = 0$ will be
\begin{align}
F(\phi) = \frac{\phi^2}{2}\frac{\partial^2F(\phi)}{\partial \phi^2} + \frac{\phi^4}{4!} \frac{\partial^4F(\phi)}{\partial \phi^4},
\label{eq:energifree}
\end{align}
because the odd terms of the expansion are null. Thus, the helicity and fourth-order helicity modulus are defined as follows:
\begin{align}
\langle \Upsilon \rangle \equiv \frac{\partial^2F(\phi)}{\partial \phi^2},
\end{align}
\begin{align}
\langle \Upsilon_4 \rangle\equiv \frac{\partial^4F(\phi)}{\partial \phi^4}.
\end{align}

As described in \cite{Minnhagen2003}, for a potential dependent solely on the angular difference between spins, the helicity modulus is expressed as:
\begin{align}
\langle \Upsilon \rangle = \langle e \rangle - N\beta\langle s^2\rangle,
\label{eq:helic}
\end{align}
with
\begin{align}
     e = \frac{1}{N}\sum_{\langle ij\rangle_x}\cos(\theta_i-\theta_j), \\
     s = \frac{1}{N}\sum_{\langle ij\rangle_x}\sin(\theta_i -\theta_j),
\end{align}
and fourth order helicity can be written as:
\begin{align}
\langle \Upsilon_4\rangle = -\frac{1}{N}\langle e\rangle +4\beta\langle s^2\rangle -3\beta[\langle \Upsilon^2\rangle-\langle \Upsilon \rangle^2] + 2\beta^3N^2\langle s^4\rangle.
\label{eq:helic4o}
\end{align}
It is important to state that equations \ref{eq:helic} and \ref{eq:helic4o} must be multiplied by the factor of $\dfrac{4}{3\sqrt{3}}$ in the case of the honeycomb lattice. This account for the ratio of spin densities when compared to the square lattice.

Therefore, $\langle \Upsilon\rangle \geq 0$. For small variations of $\phi$, the free energy is dominated by the smallest non-zero derivative, which is positive. Moreover, $\langle \Upsilon_4\rangle$ must remain positive for any temperature $T$ where $\langle \Upsilon\rangle=0$. This implies that $\langle \Upsilon\rangle$ cannot continually go to zero at the transition temperature $T_{BKT}$ if $\langle \Upsilon_4\rangle$ at the same time approaches a minimum at $T_{BKT} $. Consequently, if $\langle \Upsilon_4\rangle$ approaches the minimum in $T_{BKT}$, then the change in $\langle \Upsilon\rangle$ must be discontinuous\cite{Minnhagen2003}. Thus, $\langle \Upsilon\rangle$ must exhibit a discontinuity, supporting Minnhagen's conclusion that both helicity orders can serve as order parameters for the BKT transition\cite{Minnhagen2003}.

In order to estimate the critical BKT transition temperature, we use a second-order helicity property called the Nelson-Kosterlitz criterion\cite{Nelson_and_Kosterlitz_1977}, which predicts the following:
\begin{align}
\lim_{L\to \infty} \Upsilon(\beta_{BKT}) = \frac{2T_{BKT}}{\pi},
\label{eq:tcross}
\end{align}
This equality holds in the thermodynamic limit, as the system size draw near to infinity. The procedure involves calculating the helicity modulus for each temperature across various lattice sizes and comparing the results with the curve $r = 2T/\pi$. The intersection points determine the critical temperature for each lattice size. To estimate the BKT transition temperature using the fourth-order helicity modulus (Eq. \ref{eq:helic4o}), one simply identifies the temperature at which the function reaches its minimum.
\begin{figure}[!t]
\centering
\includegraphics[scale=0.2]{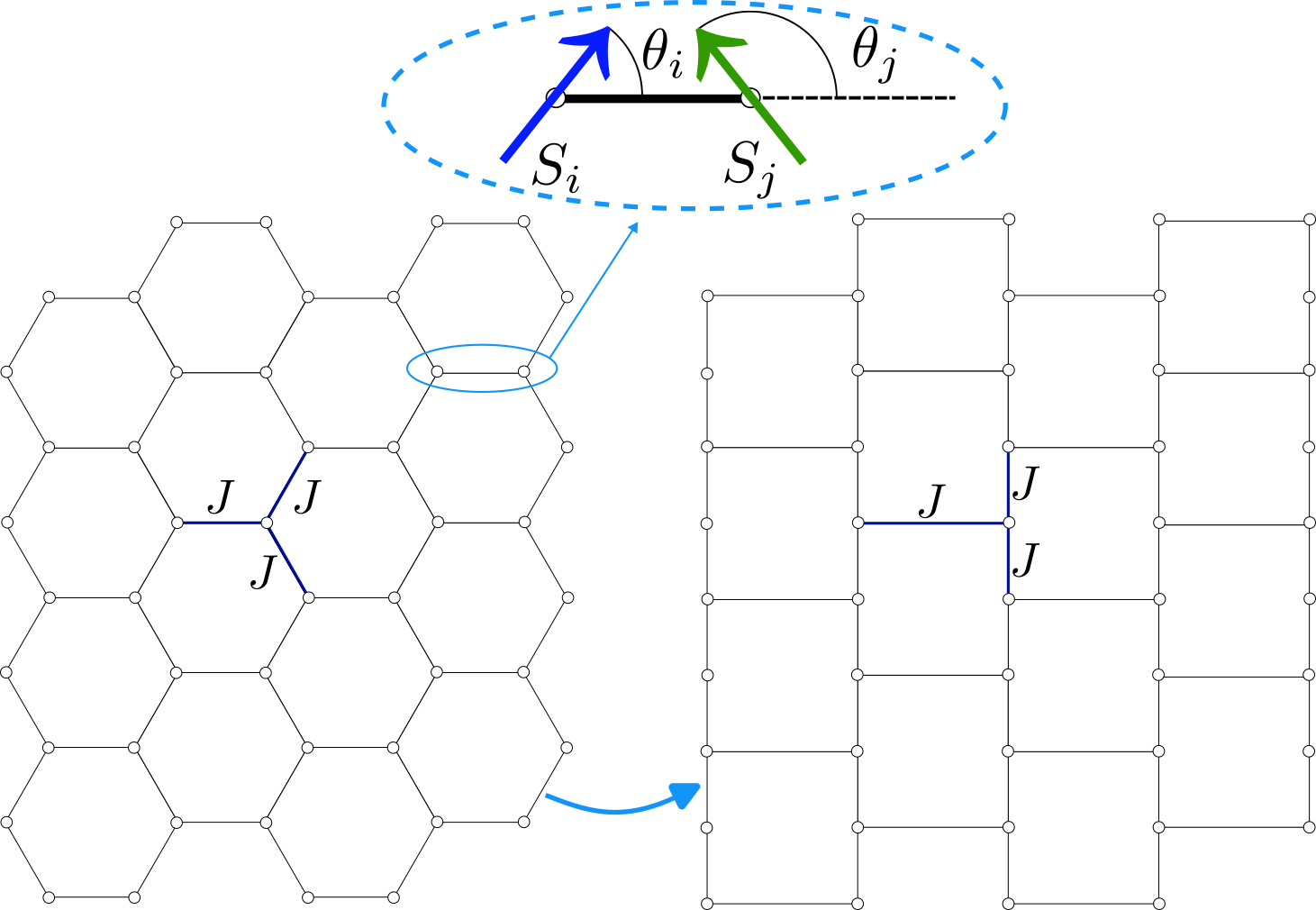}
\caption{Schematic representation of the honeycomb lattice, the spin vectors $\v{S}_i$, the first neighbors and the transposition to a square lattice.}
\label{fig:01}
\end{figure}

To address the limitation imposed by the finite size of the system in the simulation, we will employ the Finite-Size Scaling (FSS) technique, as developed by K. Binder\cite{Binder1981}. For temperatures below the critical value, some physical quantities of interest, such as helicity $\Upsilon$, can be described by a power law\cite{LandauDMXY2022}, as 
$\Upsilon \propto L^{-x}$, where $x$ is an exponent dependent on the system temperature. When the order parameter has a strong dependence on the size of the system and it is possible to find a singularity in the phase transition, a linear extrapolation ($x=1$) is appropriate to find the critical temperature in the thermodynamic limit ($L\rightarrow \infty$). Otherwise, it cannot be guaranteed that the order parameter is discontinuous in this limit. This appears to be the case for second-order helicity.  However, to mitigate this challenge, some authors propose the use of a logarithmic correction, as described in\cite{landau_binder_2014, LandauDMXY2022, Binder1997}:
\begin{align}
T_c = T_{BKT}+\frac{B}{(\ln L)^2} ,
\label{eq:fss}
\end{align}
where $T_c$ corresponds to the values of the critical transition temperature obtained by the simulations, $B$ is a dimensionless constant, and $T_{BKT}$ is the critical transition temperature in the thermodynamic limit.
\begin{figure}[!h]
\centering
\includegraphics[scale=0.3]{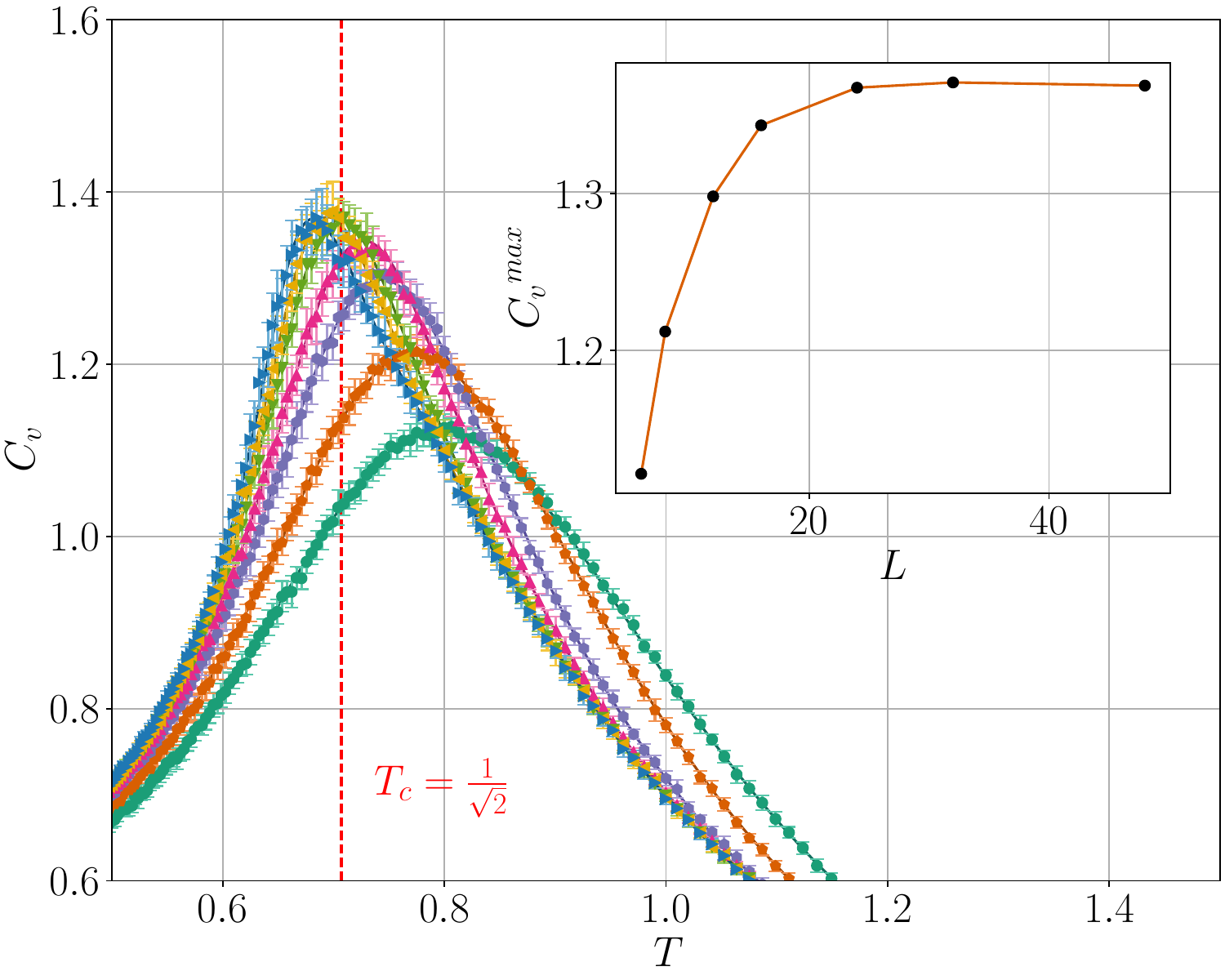}
\includegraphics[scale=0.3]{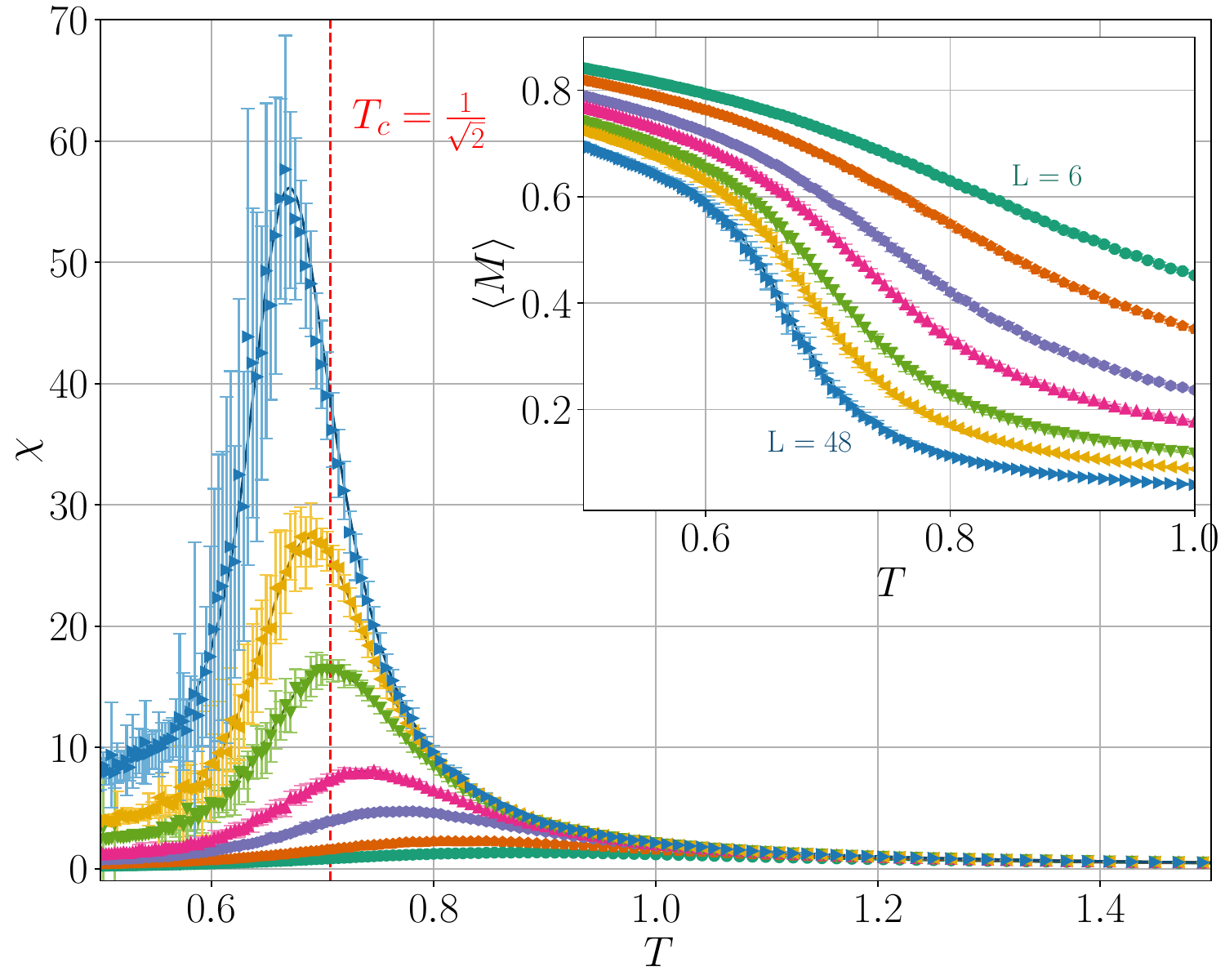}
\caption{(top) Specific heat with its maximum value as a function of $L$ in the inset and (bottom) magnetic susceptibility with average magnetization in the inset for different lattices sizes. The dots correspond to the SA results and the full line to the WL ones. The vertical dashed line indicates the theoretical expected value for $T_c$.}
\label{Medias_Flut}
\end{figure}

\section{Results}

Figure \ref{fig:01} illustrates the honeycomb lattice and its transposition to the equivalent square lattice used in the simulations. This transposition significantly simplifies the implementation of the numerical algorithm in the MC simulations without any loss of generality.

All SA simulations were conducted on lattices with sizes ranging from $L = 6$ to $L = 128$ with periodic boundary conditions. For the SA protocol, we set the initial temperature to $\beta_0 = 0.1$ and the final temperature to $\beta_n = 2$, with an increment of $\Delta\beta = 0.01$. This involved a total of $10^5$ MC sweeps for thermalization part and the same amount for acquire the thermal averages. Additionally, 20 independent simulations were performed for each lattice size to compute statistical errors via the bootstrap method. For the largest lattice size, all spin configurations were saved at each temperature to analyze the structural order of the spins and calculate the vortex density.
\begin{figure}[!t]
\centering
\includegraphics[scale=0.33]{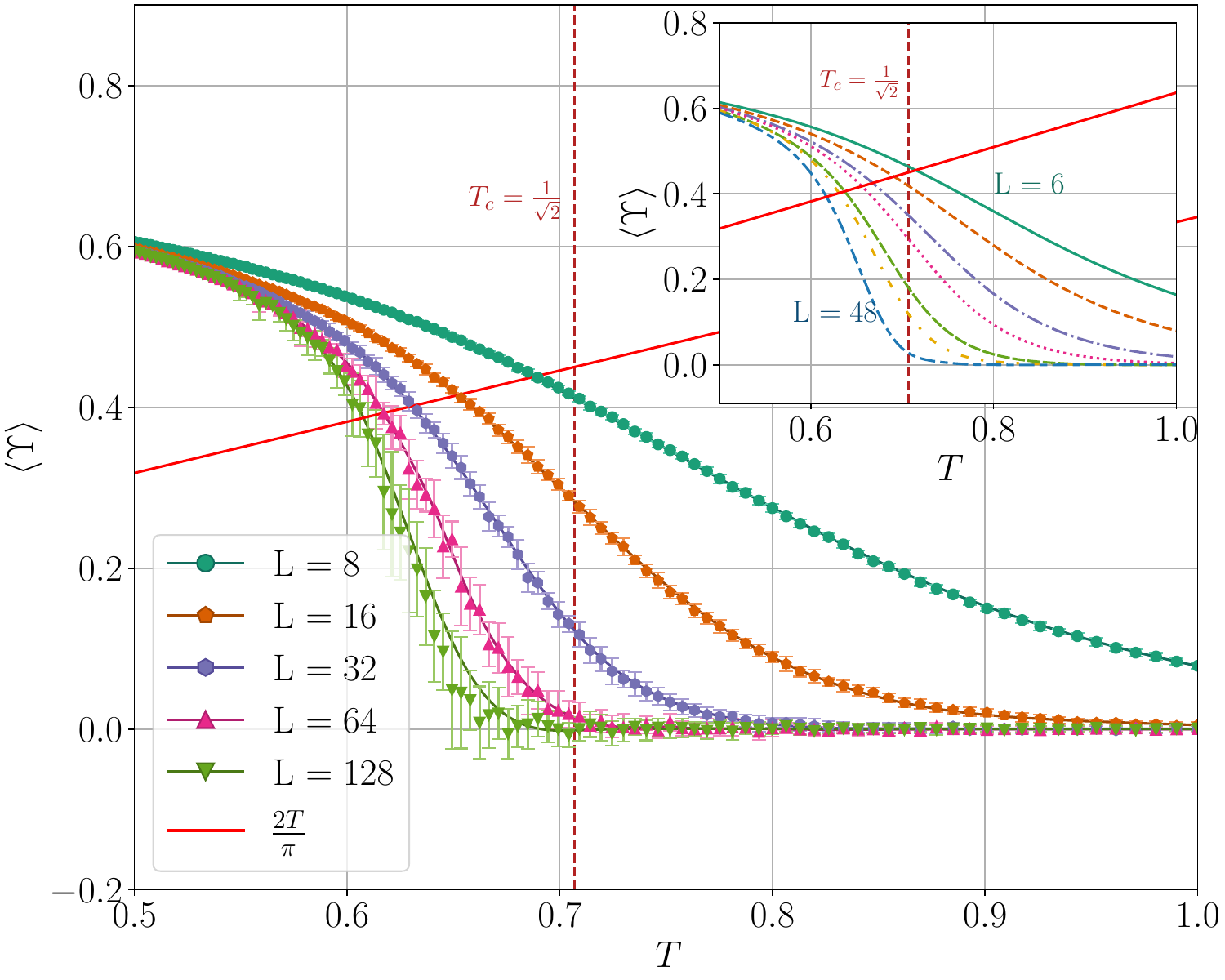}
\caption{Average second-order helicity as a function of inverse temperature for different lattice sizes obtained by SA. The inset shows the same result obtained by WL. The theoretical solid curve $2T/\pi$ indicates the critical temperature through the intersection with the helicity curve. The vertical dashed line indicates the theoretical expected value for $T_c$.\label{helic}}
\end{figure}

In the WL procedure, we adopted a similar approach to that used in Ref. \cite{Xu2007} for calculating the density of states in continuous models. Since the boundaries of the density of states spectrum exhibit large fluctuations, the flatness criterion requires a significant amount of computational time. To manage this limitation, we considered the entire energy range, which is $3N$ (where $N = L \times L$), and divided it into $N^2/4$ energy levels (bins). The standard flatness criterion is applied only to the middle spectrum range, specifically $[-N^2/4 + N, N^2/4 - N]$. For the boundary levels, we used an 80\% flatness criterion. This approach significantly reduces computational time.

Figure \ref{Medias_Flut} presents various thermodynamic quantities as a function of temperature for both SA (indicated by dots) and WL (represented by solid lines) procedures. Since the WL simulation is constrained to smaller system sizes, we present results only for $L$ values up to $48$. It is worth-noting the high resolution of the WL results when compared to the SA ones. The specific heat value approaches $1/2$ as $T \to 0$, regardless the lattice size, which aligns with the expectations of the equipartition theorem. Although both the specific heat and susceptibility exhibit a peak slightly above $T = 0.6$, these peaks are not used to determine the BKT transition temperature due to their lack of accuracy. Additionally, the peak of the specific heat does not display a consistent behavior as the system size increases (as can be seen in the inset of the same figure).
\begin{figure}[!t]
\centering
\includegraphics[scale=0.33]{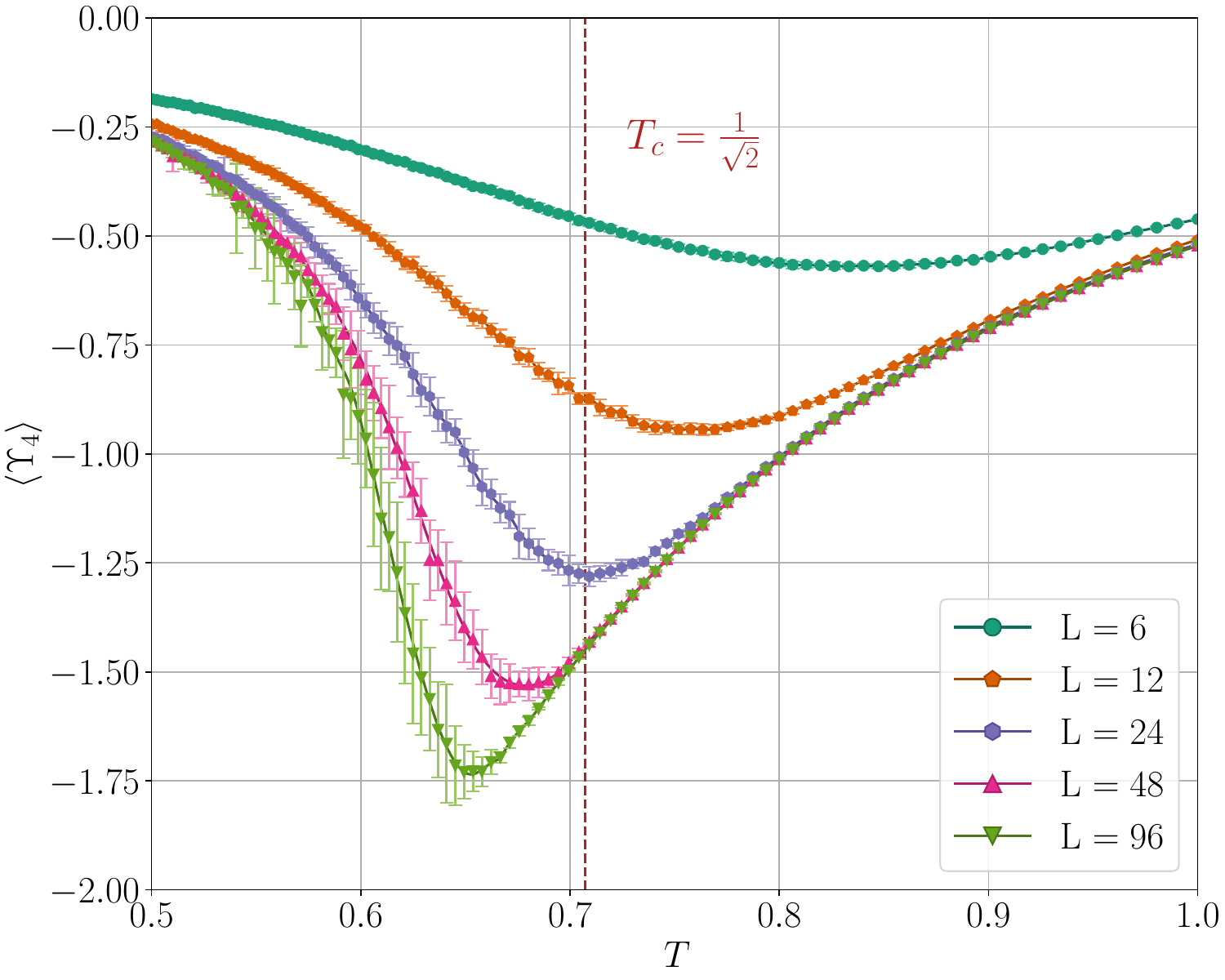}
\caption{Average fourth-order helicity modulus as a function of the inverse temperature for various lattice sizes. The vertical dashed line indicates the theoretical expected value for $T_c$.\label{helic4o}}
\end{figure}

For a precise determination of the critical temperature, we use the helicity modulus obtained from both SA and WL procedures as a function of the temperature for various lattice sizes. Figure \ref{helic} shows some of these results for SA and the inset shows the results for WL up to lattice $L=48$. The theoretical curve, represented by $2T/\pi$, is also shown. The discontinuity of the helicity modulus near the transition is not very pronounced. As a discontinuity occurs in the thermodynamic limit, we can use the intersection of the theoretical curve $2T/\pi$ with the simulated values to estimate the critical temperature.  As the lattice size increases, the intersection point shifts to the right, converging toward the critical temperature in the thermodynamic limit. A key challenge lies in accurately identifying this intersection, as the SA data are discrete. Moreover, obtaining a larger sample size with SA would incur significant computational costs. This issue does not arise with WL simulations, but they are limited as system size increases.
\begin{figure}[!t]
\centering
\includegraphics[scale=0.5]{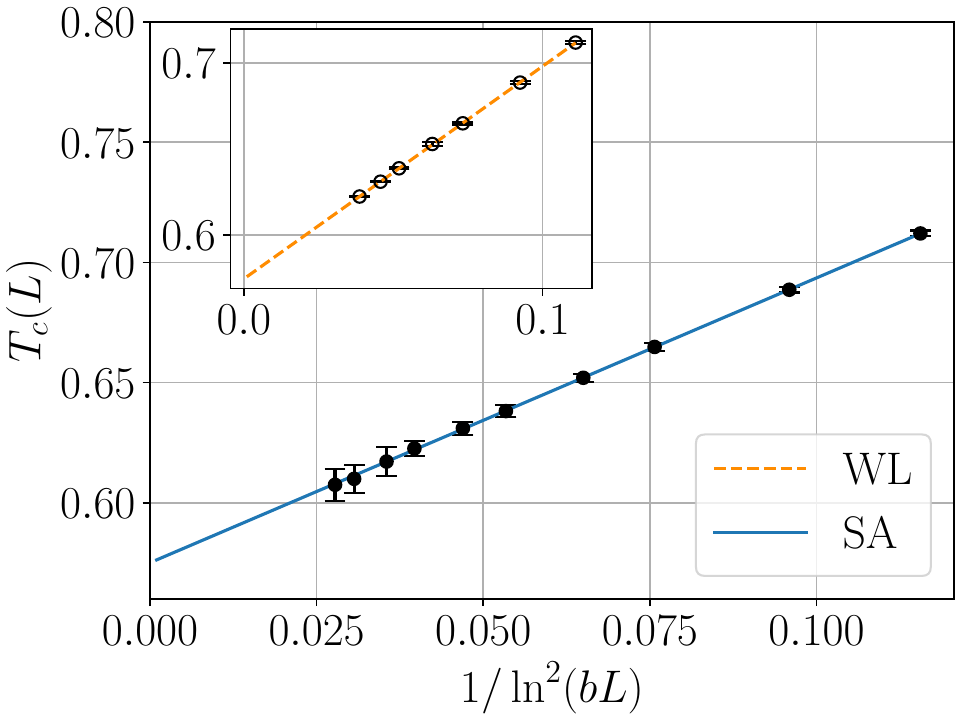}
\includegraphics[scale=0.5]{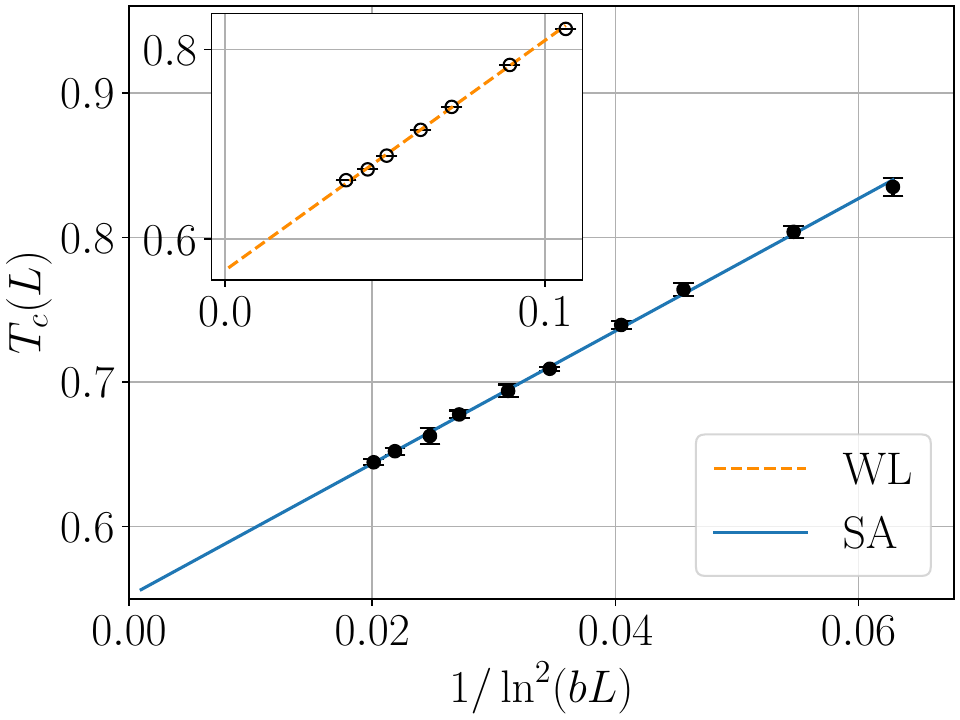}
\caption{Finite-size scaling for both $\langle \Upsilon\rangle$ (top) and $\langle \Upsilon_4\rangle$ (bottom) obtained via 
 SA and WL approaches.\label{FSS_H}}
\end{figure}

The same procedure was used to obtain fourth-order helicity modulus as a function of temperature, as shown in Fig. \ref{helic4o} for various lattice sizes. As shown by Minnhagen\cite{Minnhagen2003}, this quantity has a minimum in the BKT transition and this minimum clearly presents monotonic behavior as the system size increases. It is safe to associate this tendency with a divergent behavior as $L\to\infty$ and the corresponding temperature as the critical one. 

Based on this, we individually calculated the values of $T_c$ for each lattice and performed a finite-size scaling analysis using an improved method proposed by Hsieh et al.\cite{Hsieh_2013}. We used one of their forms for the scaling relation, i.e.
\begin{align}
    T_c(L)=T_{BKT}(\infty)+\frac{a}{\ln^2(bL)},
    \label{eq:hsieh}
\end{align}
for all simulated $T_c$. Figure \ref{FSS_H} (top) shows our FSS results for both $\langle \Upsilon\rangle$ and $\langle \Upsilon_4\rangle$. Using the helicity modulus, we obtained $T_{BKT} = 0.575(8)$ via SA and $T_{BKT}=0.576(3)$ via WL.  It is important to note that this value is significantly lower than the theoretical prediction of $1/\sqrt{2}$. Although the WL technique is limited to the system size, the result is highly accurate. This is not a surprise since the logarithmic correction proposed by Weber and Minnhagen\cite{Weber1988} and improved by Hsieh et al. to the study of the BKT transition is supposed to give accurate results. This agrees with the empirical observation that Equation \ref{eq:helic} is valid to extremely good approximation even for small lattice sizes. Very recently, F. Jiang reported neural network (NN) calculations of the same problem\cite{Jiang2024} and found a value of $0.572(3)$ for $T_{BKT}$, which is in very good agreement with our findings.

Obtaining the critical temperature through \(\langle \Upsilon_4 \rangle\) is relatively straightforward. We calculated $T_c$ for each lattice size and performed the FSS to determine $T_{BKT}$, as shown in Fig. \ref{FSS_H} (bottom). We found $T_{BKT} = 0.551(11)$ using the SA method, and $T_{BKT} = 0.568(1)$ from the WL. It is important to note that the values obtained from $\langle \Upsilon \rangle$ and $\langle \Upsilon_4 \rangle$ using SA do not agree within their error margins. This discrepancy may arise from the gradual approach of the minimum of $\langle \Upsilon_4 \rangle$ toward the critical temperature in the thermodynamic limit, suggesting that the scaling relation presented in Eq. \ref{eq:hsieh} may not be appropriate in this context. Another challenge is that the data obtained via SA near the minimum exhibit significant fluctuations, which can lead to an inaccurate determination of $T_c$. In contrast, the data from the WL method allow for a higher-resolution search for the corresponding critical temperature. However, there is still a slight discrepancy when compared to the $T_{BKT}$ obtained using the helicity modulus but agrees within the error bar with the NN calculations. The FSS analysis using $\langle \Upsilon_4 \rangle$ may require all logarithmic corrections to be taken into account in a more precise investigation of the BKT transition. Nevertheless, our results still suggest that the theoretical predictions are not fully supported by the numerical analysis. The close form of $T_c$ for $O(N)$ vector models proposed by Nienhuis\cite{Nienhuis1982} may not capture all the universal picture of this kind of model.
\begin{figure}[!t]
\centering
\includegraphics[scale=0.42]{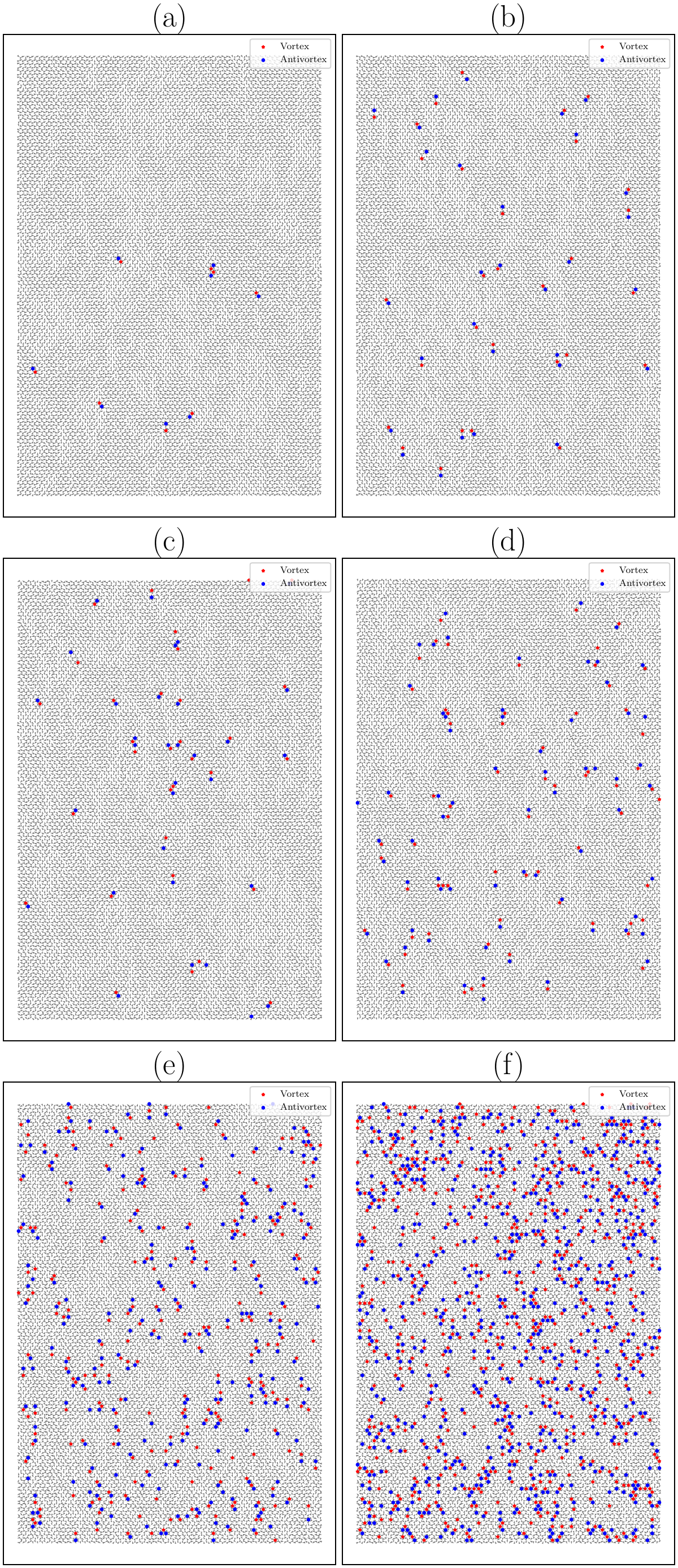}
\caption{Spin configurations for lattice size $L = 128$ and some values of temperatures: (a) $T = 0.500$, (b) $T = 0.549$, (c) $T = 0.575$, (d) $T = 0.60$ , (e) $T = 0.71$ and (f) $T = 0.91$. The red stars (blue dots) indicate the location of a vortex (antivortex).}
\label{fig:config}
\end{figure}

It is helpful to illustrate the topology of spin configurations at different temperatures as the BKT transition passes through. Figure \ref{fig:config} displays six configurations for the lattice size of $L = 128$. The red stars (blue dots) indicate the location of a vortex (antivortex). For the temperature $T=0.50$, below the critical temperature, one can see some vortex-antivortex pairs bound. The vortex density is very low, ideally tending to zero in the thermodynamic limit.
The density rises very slowly with $T$. For $T\approx0.57$, in the region where the transition occurs, there is a sudden increase in vortex density as vortex-antivortex pairs unbind. This is not a sharp jump (as in first-order phase transitions), but a rapid crossover. As the temperature rises above $T_{BKT}$, vortices and antivortices proliferate, leading to an increase in disorder until complete dissociation occurs. At this point, the system reaches a saturation point in the number of vortices.

\begin{figure}[!t]
\centering
\includegraphics[scale=0.5]{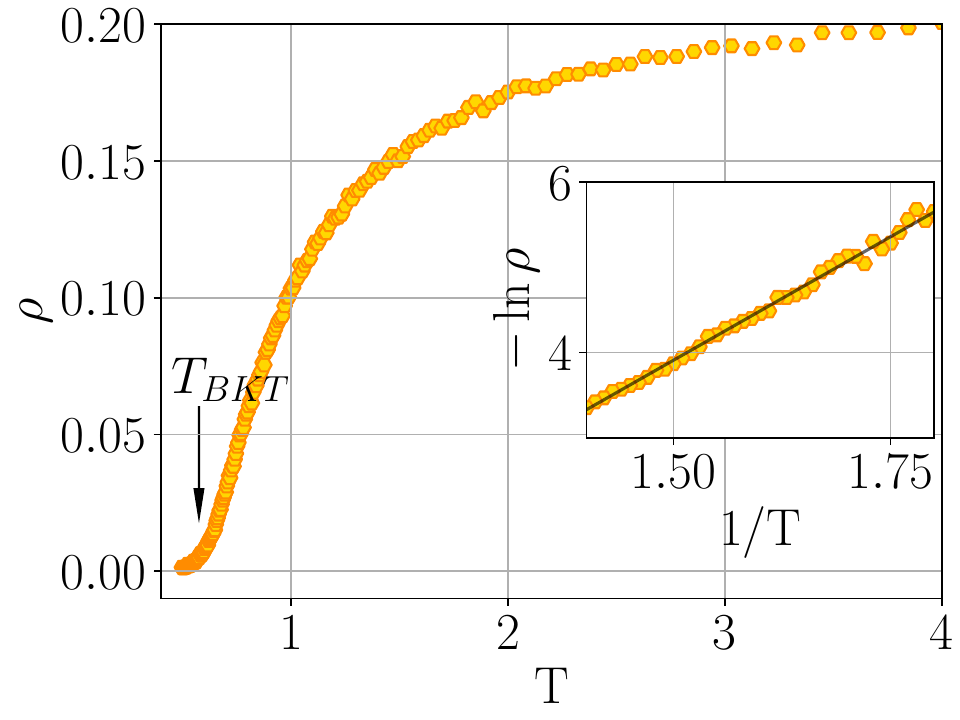}
\caption{Vortex density as a function of temperature for $L=128$. The inset shows the linear fitting given by Eq. \ref{eq:fitmu}.}
\label{fig:vortex}
\end{figure}

To perform a quantitative analysis of the dissociation of vortex-antivortex pairs, we can calculate the vortex density. A vortex is identified when the sum of \(\theta_i - \theta_j\) along a closed path at each lattice site is a multiple of \(\pm 2\pi\). Using this criterion, we determine the vortex density for various temperatures within our simulated range on a lattice of size \(L=128\), where the vortex density is defined as the number of vortices divided by \(L^2\). Figure \ref{fig:vortex} illustrates how the vortex density varies as a function of temperature. As anticipated, the density increases with higher temperatures\cite{Sun2022}. It is also noteworthy that the rate of increase in density is more pronounced near the BKT transition temperature, which occurs at a value significantly lower than the theoretical prediction. This observation supports our FSS analysis.

Finally, we can calculate the formation energy of a pair of vortices. Near the transition temperature, the vortex density behavior follows the relationship\cite{Sun2022}
\begin{align}
     \rho \sim \exp (-2\mu/T),
\end{align}
where $2\mu$ represents the formation energy of a pair of vortex. By applying some algebra, we can derive the following linear relationship between $-\ln(\rho)$ and $1/T$
\begin{align}
     -\ln(\rho) \sim 2\mu/T,
     \label{eq:fitmu}
\end{align}
where $2\mu$ is the angular coefficient of this function. Using this relationship, we adjusted the vortex density values close to the critical temperature and obtained $2\mu = 5.80(12)$ in units of $J$. This value is lower than the one obtained for the square lattice\cite{gupta1992}. This result is justified by the idea of the instability of the honeycomb lattice to support the spin long-range order, as investigated by Wojtkiewicz et al.\cite{Wojtkiewicz2023}.

\section{Conclusions}

We conducted a thorough analysis of the XY model on a honeycomb lattice using the numerical Monte Carlo method, employing two techniques: the Simulated Annealing protocol and high-resolution entropic simulations based on the Wang-Landau algorithm. Our primary goal was to calculate the critical transition temperature and to compare our results with theoretical predictions.

To achieve this, we utilized the second- and fourth-order helicity modulus as order parameters, as these properties exhibit a discontinuity at the BKT phase transition. Our best estimates for the transition temperature were obtained using the helicity modulus from both techniques: we found $T_{BKT} = 0.575(8)$ from the Simulated Annealing method and $T_{BKT} = 0.576(3)$ from the Wang-Landau algorithm.

Furthermore, we assessed the applicability of the fourth-order helicity modulus in determining the BKT temperature. While it presents a minimum at the transition that scales with system size, the large fluctuations in the data obtained from Simulated Annealing hinder the precise identification of the critical temperature. In contrast, the data from the Wang-Landau method is more precise and smoother, although it is limited to larger system sizes, yielding values of $T_{BKT} = 0.551(11)$ from Simulated Annealing and $T_{BKT} = 0.568(1)$ from Wang-Landau.

Interestingly, these results align closely with those obtained independently through neural network calculations. Notably, these values are significantly lower than the analytical prediction of $1/\sqrt{2}$. Our findings suggest that theoretical calculations of the critical temperature for the XY model on the honeycomb lattice need to be reevaluated.

Additionally, we calculated the vortex density, which allowed us to estimate the formation energy of a pair of vortices. The obtained value was $2\mu = 5.80(12)$ in units of $J$, which is smaller than the corresponding value for the square lattice, justified by the idea of instability of the honeycomb lattice to support the spin long-range order.

\bibliography{references}

\begin{thebibliography}{39}%
\makeatletter
\providecommand \@ifxundefined [1]{%
 \@ifx{#1\undefined}
}%
\providecommand \@ifnum [1]{%
 \ifnum #1\expandafter \@firstoftwo
 \else \expandafter \@secondoftwo
 \fi
}%
\providecommand \@ifx [1]{%
 \ifx #1\expandafter \@firstoftwo
 \else \expandafter \@secondoftwo
 \fi
}%
\providecommand \natexlab [1]{#1}%
\providecommand \enquote  [1]{``#1''}%
\providecommand \bibnamefont  [1]{#1}%
\providecommand \bibfnamefont [1]{#1}%
\providecommand \citenamefont [1]{#1}%
\providecommand \href@noop [0]{\@secondoftwo}%
\providecommand \href [0]{\begingroup \@sanitize@url \@href}%
\providecommand \@href[1]{\@@startlink{#1}\@@href}%
\providecommand \@@href[1]{\endgroup#1\@@endlink}%
\providecommand \@sanitize@url [0]{\catcode `\\12\catcode `\$12\catcode `\&12\catcode `\#12\catcode `\^12\catcode `\_12\catcode `\%12\relax}%
\providecommand \@@startlink[1]{}%
\providecommand \@@endlink[0]{}%
\providecommand \url  [0]{\begingroup\@sanitize@url \@url }%
\providecommand \@url [1]{\endgroup\@href {#1}{\urlprefix }}%
\providecommand \urlprefix  [0]{URL }%
\providecommand \Eprint [0]{\href }%
\providecommand \doibase [0]{http://dx.doi.org/}%
\providecommand \selectlanguage [0]{\@gobble}%
\providecommand \bibinfo  [0]{\@secondoftwo}%
\providecommand \bibfield  [0]{\@secondoftwo}%
\providecommand \translation [1]{[#1]}%
\providecommand \BibitemOpen [0]{}%
\providecommand \bibitemStop [0]{}%
\providecommand \bibitemNoStop [0]{.\EOS\space}%
\providecommand \EOS [0]{\spacefactor3000\relax}%
\providecommand \BibitemShut  [1]{\csname bibitem#1\endcsname}%
\let\auto@bib@innerbib\@empty
\bibitem [{\citenamefont {Berezinskii}(1972)}]{Berezinski1972}%
  \BibitemOpen
  \bibfield  {author} {\bibinfo {author} {\bibfnamefont {V.~L.}\ \bibnamefont {Berezinskii}},\ }\href@noop {} {\bibfield  {journal} {\bibinfo  {journal} {Soviet Journal of Experimental and Theoretical Physics}\ }\textbf {\bibinfo {volume} {34}},\ \bibinfo {pages} {610} (\bibinfo {year} {1972})}\BibitemShut {NoStop}%
\bibitem [{\citenamefont {Kosterlitz}\ and\ \citenamefont {Thouless}(1973)}]{Kosterlitz1973}%
  \BibitemOpen
  \bibfield  {author} {\bibinfo {author} {\bibfnamefont {J.~M.}\ \bibnamefont {Kosterlitz}}\ and\ \bibinfo {author} {\bibfnamefont {D.~J.}\ \bibnamefont {Thouless}},\ }\href@noop {} {\bibfield  {journal} {\bibinfo  {journal} {Journal of Physics C: Solid State Physics}\ }\textbf {\bibinfo {volume} {6}},\ \bibinfo {pages} {1181} (\bibinfo {year} {1973})}\BibitemShut {NoStop}%
\bibitem [{\citenamefont {Kosterlitz}(1974)}]{Kosterlitz_1974}%
  \BibitemOpen
  \bibfield  {author} {\bibinfo {author} {\bibfnamefont {J.~M.}\ \bibnamefont {Kosterlitz}},\ }\href@noop {} {\bibfield  {journal} {\bibinfo  {journal} {Journal of Physics C: Solid State Physics}\ }\textbf {\bibinfo {volume} {7}},\ \bibinfo {pages} {1046} (\bibinfo {year} {1974})}\BibitemShut {NoStop}%
\bibitem [{\citenamefont {Babaev}(2004)}]{BABAEV2004}%
  \BibitemOpen
  \bibfield  {author} {\bibinfo {author} {\bibfnamefont {E.}~\bibnamefont {Babaev}},\ }\href {\doibase https://doi.org/10.1016/j.nuclphysb.2004.02.021} {\bibfield  {journal} {\bibinfo  {journal} {Nuclear Physics B}\ }\textbf {\bibinfo {volume} {686}},\ \bibinfo {pages} {397} (\bibinfo {year} {2004})}\BibitemShut {NoStop}%
\bibitem [{\citenamefont {Hlubina}(2008)}]{Hlubina2008}%
  \BibitemOpen
  \bibfield  {author} {\bibinfo {author} {\bibfnamefont {R.}~\bibnamefont {Hlubina}},\ }\href {\doibase 10.1103/PhysRevB.77.094503} {\bibfield  {journal} {\bibinfo  {journal} {Phys. Rev. B}\ }\textbf {\bibinfo {volume} {77}},\ \bibinfo {pages} {094503} (\bibinfo {year} {2008})}\BibitemShut {NoStop}%
\bibitem [{\citenamefont {Bruinsma}\ and\ \citenamefont {Aeppli}(1982)}]{Bruinsma1982}%
  \BibitemOpen
  \bibfield  {author} {\bibinfo {author} {\bibfnamefont {R.}~\bibnamefont {Bruinsma}}\ and\ \bibinfo {author} {\bibfnamefont {G.}~\bibnamefont {Aeppli}},\ }\href {\doibase 10.1103/PhysRevLett.48.1625} {\bibfield  {journal} {\bibinfo  {journal} {Phys. Rev. Lett.}\ }\textbf {\bibinfo {volume} {48}},\ \bibinfo {pages} {1625} (\bibinfo {year} {1982})}\BibitemShut {NoStop}%
\bibitem [{\citenamefont {Aeppli}\ and\ \citenamefont {Bruinsma}(1984)}]{Aeppli1984}%
  \BibitemOpen
  \bibfield  {author} {\bibinfo {author} {\bibfnamefont {G.}~\bibnamefont {Aeppli}}\ and\ \bibinfo {author} {\bibfnamefont {R.}~\bibnamefont {Bruinsma}},\ }\href {\doibase 10.1103/PhysRevLett.53.2133} {\bibfield  {journal} {\bibinfo  {journal} {Phys. Rev. Lett.}\ }\textbf {\bibinfo {volume} {53}},\ \bibinfo {pages} {2133} (\bibinfo {year} {1984})}\BibitemShut {NoStop}%
\bibitem [{\citenamefont {Minnhagen}\ and\ \citenamefont {Kim}(2003)}]{Minnhagen2003}%
  \BibitemOpen
  \bibfield  {author} {\bibinfo {author} {\bibfnamefont {P.}~\bibnamefont {Minnhagen}}\ and\ \bibinfo {author} {\bibfnamefont {B.~J.}\ \bibnamefont {Kim}},\ }\href@noop {} {\bibfield  {journal} {\bibinfo  {journal} {Phys. Rev. B}\ }\textbf {\bibinfo {volume} {67}},\ \bibinfo {pages} {172509} (\bibinfo {year} {2003})}\BibitemShut {NoStop}%
\bibitem [{\citenamefont {Deng}\ and\ \citenamefont {Gu}(2014)}]{Deng_2014}%
  \BibitemOpen
  \bibfield  {author} {\bibinfo {author} {\bibfnamefont {Y.-B.}\ \bibnamefont {Deng}}\ and\ \bibinfo {author} {\bibfnamefont {Q.}~\bibnamefont {Gu}},\ }\href {\doibase 10.1088/0256-307x/31/2/020504} {\bibfield  {journal} {\bibinfo  {journal} {Chinese Physics Letters}\ }\textbf {\bibinfo {volume} {31}},\ \bibinfo {pages} {020504} (\bibinfo {year} {2014})}\BibitemShut {NoStop}%
\bibitem [{\citenamefont {Lach}\ and\ \citenamefont {\ifmmode \check{Z}\else \v{Z}\fi{}ukovi\ifmmode~\check{c}\else \v{c}\fi{}}(2021)}]{Lach2021}%
  \BibitemOpen
  \bibfield  {author} {\bibinfo {author} {\bibfnamefont {M.}~\bibnamefont {Lach}}\ and\ \bibinfo {author} {\bibfnamefont {M.}~\bibnamefont {\ifmmode \check{Z}\else \v{Z}\fi{}ukovi\ifmmode~\check{c}\else \v{c}\fi{}}},\ }\href {\doibase 10.1103/PhysRevE.104.024134} {\bibfield  {journal} {\bibinfo  {journal} {Phys. Rev. E}\ }\textbf {\bibinfo {volume} {104}},\ \bibinfo {pages} {024134} (\bibinfo {year} {2021})}\BibitemShut {NoStop}%
\bibitem [{\citenamefont {Sun}\ \emph {et~al.}(2022)\citenamefont {Sun}, \citenamefont {Wu}, \citenamefont {Yang}, \citenamefont {Zhou}, \citenamefont {Zhu}, \citenamefont {Chen},\ and\ \citenamefont {An}}]{Sun2022}%
  \BibitemOpen
  \bibfield  {author} {\bibinfo {author} {\bibfnamefont {Y.-Z.}\ \bibnamefont {Sun}}, \bibinfo {author} {\bibfnamefont {Q.}~\bibnamefont {Wu}}, \bibinfo {author} {\bibfnamefont {X.-L.}\ \bibnamefont {Yang}}, \bibinfo {author} {\bibfnamefont {Y.}~\bibnamefont {Zhou}}, \bibinfo {author} {\bibfnamefont {L.-Y.}\ \bibnamefont {Zhu}}, \bibinfo {author} {\bibfnamefont {Q.}~\bibnamefont {Chen}}, \ and\ \bibinfo {author} {\bibfnamefont {Q.}~\bibnamefont {An}},\ }\href {\doibase 10.3389/fphy.2022.851322} {\bibfield  {journal} {\bibinfo  {journal} {Frontiers in Physics}\ }\textbf {\bibinfo {volume} {10}} (\bibinfo {year} {2022}),\ 10.3389/fphy.2022.851322}\BibitemShut {NoStop}%
\bibitem [{\citenamefont {Silva}\ \emph {et~al.}(2022)\citenamefont {Silva}, \citenamefont {Plascak},\ and\ \citenamefont {Landau}}]{LandauDMXY2022}%
  \BibitemOpen
  \bibfield  {author} {\bibinfo {author} {\bibfnamefont {G.~A.}\ \bibnamefont {Silva}}, \bibinfo {author} {\bibfnamefont {J.~A.}\ \bibnamefont {Plascak}}, \ and\ \bibinfo {author} {\bibfnamefont {D.~P.}\ \bibnamefont {Landau}},\ }\href@noop {} {\bibfield  {journal} {\bibinfo  {journal} {Phys. Rev. E}\ }\textbf {\bibinfo {volume} {106}},\ \bibinfo {pages} {044116} (\bibinfo {year} {2022})}\BibitemShut {NoStop}%
\bibitem [{\citenamefont {Cui}\ \emph {et~al.}(2022)\citenamefont {Cui}, \citenamefont {Wang}, \citenamefont {Zhu}, \citenamefont {Liang},\ and\ \citenamefont {Yang}}]{Cui2022}%
  \BibitemOpen
  \bibfield  {author} {\bibinfo {author} {\bibfnamefont {Q.}~\bibnamefont {Cui}}, \bibinfo {author} {\bibfnamefont {L.}~\bibnamefont {Wang}}, \bibinfo {author} {\bibfnamefont {Y.}~\bibnamefont {Zhu}}, \bibinfo {author} {\bibfnamefont {J.}~\bibnamefont {Liang}}, \ and\ \bibinfo {author} {\bibfnamefont {H.}~\bibnamefont {Yang}},\ }\href {\doibase 10.1007/s11467-022-1217-7} {\bibfield  {journal} {\bibinfo  {journal} {Frontiers of Physics}\ }\textbf {\bibinfo {volume} {18}},\ \bibinfo {pages} {13602} (\bibinfo {year} {2022})}\BibitemShut {NoStop}%
\bibitem [{\citenamefont {Drouin-Touchette}\ \emph {et~al.}(2022)\citenamefont {Drouin-Touchette}, \citenamefont {Orth}, \citenamefont {Coleman}, \citenamefont {Chandra},\ and\ \citenamefont {Lubensky}}]{Touchette2022}%
  \BibitemOpen
  \bibfield  {author} {\bibinfo {author} {\bibfnamefont {V.}~\bibnamefont {Drouin-Touchette}}, \bibinfo {author} {\bibfnamefont {P.~P.}\ \bibnamefont {Orth}}, \bibinfo {author} {\bibfnamefont {P.}~\bibnamefont {Coleman}}, \bibinfo {author} {\bibfnamefont {P.}~\bibnamefont {Chandra}}, \ and\ \bibinfo {author} {\bibfnamefont {T.~C.}\ \bibnamefont {Lubensky}},\ }\href {\doibase 10.1103/PhysRevX.12.011043} {\bibfield  {journal} {\bibinfo  {journal} {Phys. Rev. X}\ }\textbf {\bibinfo {volume} {12}},\ \bibinfo {pages} {011043} (\bibinfo {year} {2022})}\BibitemShut {NoStop}%
\bibitem [{\citenamefont {Hsieh}\ \emph {et~al.}(2013)\citenamefont {Hsieh}, \citenamefont {Kao},\ and\ \citenamefont {Sandvik}}]{Hsieh_2013}%
  \BibitemOpen
  \bibfield  {author} {\bibinfo {author} {\bibfnamefont {Y.-D.}\ \bibnamefont {Hsieh}}, \bibinfo {author} {\bibfnamefont {Y.-J.}\ \bibnamefont {Kao}}, \ and\ \bibinfo {author} {\bibfnamefont {A.~W.}\ \bibnamefont {Sandvik}},\ }\href@noop {} {\bibfield  {journal} {\bibinfo  {journal} {Journal of Statistical Mechanics: Theory and Experiment}\ }\textbf {\bibinfo {volume} {2013}},\ \bibinfo {pages} {P09001} (\bibinfo {year} {2013})}\BibitemShut {NoStop}%
\bibitem [{\citenamefont {Nienhuis}(1982)}]{Nienhuis1982}%
  \BibitemOpen
  \bibfield  {author} {\bibinfo {author} {\bibfnamefont {B.}~\bibnamefont {Nienhuis}},\ }\href {\doibase 10.1103/PhysRevLett.49.1062} {\bibfield  {journal} {\bibinfo  {journal} {Phys. Rev. Lett.}\ }\textbf {\bibinfo {volume} {49}},\ \bibinfo {pages} {1062} (\bibinfo {year} {1982})}\BibitemShut {NoStop}%
\bibitem [{\citenamefont {Campostrini}\ \emph {et~al.}(1996)\citenamefont {Campostrini}, \citenamefont {Pelissetto}, \citenamefont {Rossi},\ and\ \citenamefont {Vicari}}]{campostrini1996}%
  \BibitemOpen
  \bibfield  {author} {\bibinfo {author} {\bibfnamefont {M.}~\bibnamefont {Campostrini}}, \bibinfo {author} {\bibfnamefont {A.}~\bibnamefont {Pelissetto}}, \bibinfo {author} {\bibfnamefont {P.}~\bibnamefont {Rossi}}, \ and\ \bibinfo {author} {\bibfnamefont {E.}~\bibnamefont {Vicari}},\ }\href {\doibase 10.1103/PhysRevB.54.7301} {\bibfield  {journal} {\bibinfo  {journal} {Phys. Rev. B}\ }\textbf {\bibinfo {volume} {54}},\ \bibinfo {pages} {7301} (\bibinfo {year} {1996})}\BibitemShut {NoStop}%
\bibitem [{\citenamefont {Deng}\ \emph {et~al.}(2007)\citenamefont {Deng}, \citenamefont {Garoni}, \citenamefont {Guo}, \citenamefont {Bl\"ote},\ and\ \citenamefont {Sokal}}]{Deng2007}%
  \BibitemOpen
  \bibfield  {author} {\bibinfo {author} {\bibfnamefont {Y.}~\bibnamefont {Deng}}, \bibinfo {author} {\bibfnamefont {T.~M.}\ \bibnamefont {Garoni}}, \bibinfo {author} {\bibfnamefont {W.}~\bibnamefont {Guo}}, \bibinfo {author} {\bibfnamefont {H.~W.~J.}\ \bibnamefont {Bl\"ote}}, \ and\ \bibinfo {author} {\bibfnamefont {A.~D.}\ \bibnamefont {Sokal}},\ }\href {\doibase 10.1103/PhysRevLett.98.120601} {\bibfield  {journal} {\bibinfo  {journal} {Phys. Rev. Lett.}\ }\textbf {\bibinfo {volume} {98}},\ \bibinfo {pages} {120601} (\bibinfo {year} {2007})}\BibitemShut {NoStop}%
\bibitem [{\citenamefont {Wang}\ \emph {et~al.}(2021)\citenamefont {Wang}, \citenamefont {Zhang}, \citenamefont {Hua},\ and\ \citenamefont {Wei}}]{jielin2021}%
  \BibitemOpen
  \bibfield  {author} {\bibinfo {author} {\bibfnamefont {J.}~\bibnamefont {Wang}}, \bibinfo {author} {\bibfnamefont {W.}~\bibnamefont {Zhang}}, \bibinfo {author} {\bibfnamefont {T.}~\bibnamefont {Hua}}, \ and\ \bibinfo {author} {\bibfnamefont {T.-C.}\ \bibnamefont {Wei}},\ }\href {\doibase 10.1103/PhysRevResearch.3.013074} {\bibfield  {journal} {\bibinfo  {journal} {Phys. Rev. Res.}\ }\textbf {\bibinfo {volume} {3}},\ \bibinfo {pages} {013074} (\bibinfo {year} {2021})}\BibitemShut {NoStop}%
\bibitem [{\citenamefont {Caparica}(2014)}]{Caparica2014a}%
  \BibitemOpen
  \bibfield  {author} {\bibinfo {author} {\bibfnamefont {A.}~\bibnamefont {Caparica}},\ }\href@noop {} {\bibfield  {journal} {\bibinfo  {journal} {Physical Review E}\ }\textbf {\bibinfo {volume} {89}},\ \bibinfo {pages} {043301} (\bibinfo {year} {2014})}\BibitemShut {NoStop}%
\bibitem [{\citenamefont {Landau}\ and\ \citenamefont {Binder}(2014)}]{landau_binder_2014}%
  \BibitemOpen
  \bibfield  {author} {\bibinfo {author} {\bibfnamefont {D.~P.}\ \bibnamefont {Landau}}\ and\ \bibinfo {author} {\bibfnamefont {K.}~\bibnamefont {Binder}},\ }\href {\doibase 10.1017/CBO9781139696463} {\emph {\bibinfo {title} {A Guide to Monte Carlo Simulations in Statistical Physics}}},\ \bibinfo {edition} {4th}\ ed.\ (\bibinfo  {publisher} {Cambridge University Press},\ \bibinfo {year} {2014})\BibitemShut {NoStop}%
\bibitem [{\citenamefont {Wang}\ and\ \citenamefont {Landau}(2001{\natexlab{a}})}]{Wang2001a}%
  \BibitemOpen
  \bibfield  {author} {\bibinfo {author} {\bibfnamefont {F.}~\bibnamefont {Wang}}\ and\ \bibinfo {author} {\bibfnamefont {D.}~\bibnamefont {Landau}},\ }\href@noop {} {\bibfield  {journal} {\bibinfo  {journal} {Phys. Rev. Lett.}\ }\textbf {\bibinfo {volume} {86}},\ \bibinfo {pages} {2050} (\bibinfo {year} {2001}{\natexlab{a}})}\BibitemShut {NoStop}%
\bibitem [{\citenamefont {Gong}\ and\ \citenamefont {Zhang}(2019)}]{Gong2019}%
  \BibitemOpen
  \bibfield  {author} {\bibinfo {author} {\bibfnamefont {C.}~\bibnamefont {Gong}}\ and\ \bibinfo {author} {\bibfnamefont {X.}~\bibnamefont {Zhang}},\ }\href {\doibase 10.1126/science.aav4450} {\bibfield  {journal} {\bibinfo  {journal} {Science}\ }\textbf {\bibinfo {volume} {363}},\ \bibinfo {pages} {eaav4450} (\bibinfo {year} {2019})}\BibitemShut {NoStop}%
\bibitem [{\citenamefont {Gong}\ \emph {et~al.}(2017)\citenamefont {Gong}, \citenamefont {Li}, \citenamefont {Li}, \citenamefont {Ji}, \citenamefont {Stern}, \citenamefont {Xia}, \citenamefont {Cao}, \citenamefont {Bao}, \citenamefont {Wang}, \citenamefont {Wang}, \citenamefont {Qiu}, \citenamefont {Cava}, \citenamefont {Louie}, \citenamefont {Xia},\ and\ \citenamefont {Zhang}}]{Gong2017}%
  \BibitemOpen
  \bibfield  {author} {\bibinfo {author} {\bibfnamefont {C.}~\bibnamefont {Gong}}, \bibinfo {author} {\bibfnamefont {L.}~\bibnamefont {Li}}, \bibinfo {author} {\bibfnamefont {Z.}~\bibnamefont {Li}}, \bibinfo {author} {\bibfnamefont {H.}~\bibnamefont {Ji}}, \bibinfo {author} {\bibfnamefont {A.}~\bibnamefont {Stern}}, \bibinfo {author} {\bibfnamefont {Y.}~\bibnamefont {Xia}}, \bibinfo {author} {\bibfnamefont {T.}~\bibnamefont {Cao}}, \bibinfo {author} {\bibfnamefont {W.}~\bibnamefont {Bao}}, \bibinfo {author} {\bibfnamefont {C.}~\bibnamefont {Wang}}, \bibinfo {author} {\bibfnamefont {Y.}~\bibnamefont {Wang}}, \bibinfo {author} {\bibfnamefont {Z.~Q.}\ \bibnamefont {Qiu}}, \bibinfo {author} {\bibfnamefont {R.~J.}\ \bibnamefont {Cava}}, \bibinfo {author} {\bibfnamefont {S.~G.}\ \bibnamefont {Louie}}, \bibinfo {author} {\bibfnamefont {J.}~\bibnamefont {Xia}}, \ and\ \bibinfo {author} {\bibfnamefont {X.}~\bibnamefont {Zhang}},\ }\href {\doibase 10.1038/nature22060} {\bibfield  {journal} {\bibinfo  {journal}
  {Nature}\ }\textbf {\bibinfo {volume} {546}},\ \bibinfo {pages} {265} (\bibinfo {year} {2017})}\BibitemShut {NoStop}%
\bibitem [{\citenamefont {Stanley}(1971)}]{stanley1971phase}%
  \BibitemOpen
  \bibfield  {author} {\bibinfo {author} {\bibfnamefont {H.~E.}\ \bibnamefont {Stanley}},\ }\href@noop {} {\bibfield  {journal} {\bibinfo  {journal} {Clarendon, Oxford}\ ,\ \bibinfo {pages} {9}} (\bibinfo {year} {1971})}\BibitemShut {NoStop}%
\bibitem [{\citenamefont {Newman}(2012)}]{NewmanCP2012}%
  \BibitemOpen
  \bibfield  {author} {\bibinfo {author} {\bibfnamefont {M.}~\bibnamefont {Newman}},\ }\href@noop {} {\emph {\bibinfo {title} {Computational Physics}}}\ (\bibinfo  {publisher} {CreateSpace Independent Publishing Platform},\ \bibinfo {year} {2012})\BibitemShut {NoStop}%
\bibitem [{\citenamefont {Caparica}\ and\ \citenamefont {DaSilva}(2015)}]{Caparica2015}%
  \BibitemOpen
  \bibfield  {author} {\bibinfo {author} {\bibfnamefont {A.~A.}\ \bibnamefont {Caparica}}\ and\ \bibinfo {author} {\bibfnamefont {C.~J.}\ \bibnamefont {DaSilva}},\ }\href {\doibase 10.1007/s13538-015-0361-8} {\bibfield  {journal} {\bibinfo  {journal} {Brazilian Journal of Physics}\ }\textbf {\bibinfo {volume} {45}},\ \bibinfo {pages} {713} (\bibinfo {year} {2015})}\BibitemShut {NoStop}%
\bibitem [{\citenamefont {Caparica}\ and\ \citenamefont {Cunha{-}Netto}(2012)}]{Caparica2012}%
  \BibitemOpen
  \bibfield  {author} {\bibinfo {author} {\bibfnamefont {A.~A.}\ \bibnamefont {Caparica}}\ and\ \bibinfo {author} {\bibfnamefont {A.~G.}\ \bibnamefont {Cunha{-}Netto}},\ }\href {https://link.aps.org/doi/10.1103/PhysRevE.85.046702} {\bibfield  {journal} {\bibinfo  {journal} {Phys. Rev. E}\ }\textbf {\bibinfo {volume} {85}},\ \bibinfo {pages} {046702} (\bibinfo {year} {2012})}\BibitemShut {NoStop}%
\bibitem [{\citenamefont {Wang}\ and\ \citenamefont {Landau}(2001{\natexlab{b}})}]{Wang2001}%
  \BibitemOpen
  \bibfield  {author} {\bibinfo {author} {\bibfnamefont {F.}~\bibnamefont {Wang}}\ and\ \bibinfo {author} {\bibfnamefont {D.}~\bibnamefont {Landau}},\ }\href@noop {} {\bibfield  {journal} {\bibinfo  {journal} {Physical Review E}\ }\textbf {\bibinfo {volume} {64}},\ \bibinfo {pages} {056101} (\bibinfo {year} {2001}{\natexlab{b}})}\BibitemShut {NoStop}%
\bibitem [{\citenamefont {Ferreira}\ \emph {et~al.}(2018)\citenamefont {Ferreira}, \citenamefont {Jorge}, \citenamefont {Le{\~a}o},\ and\ \citenamefont {Caparica}}]{ferreira2018}%
  \BibitemOpen
  \bibfield  {author} {\bibinfo {author} {\bibfnamefont {L.}~\bibnamefont {Ferreira}}, \bibinfo {author} {\bibfnamefont {L.}~\bibnamefont {Jorge}}, \bibinfo {author} {\bibfnamefont {S.}~\bibnamefont {Le{\~a}o}}, \ and\ \bibinfo {author} {\bibfnamefont {A.}~\bibnamefont {Caparica}},\ }\href@noop {} {\bibfield  {journal} {\bibinfo  {journal} {Journal of Computational Physics}\ }\textbf {\bibinfo {volume} {358}},\ \bibinfo {pages} {130} (\bibinfo {year} {2018})}\BibitemShut {NoStop}%
\bibitem [{\citenamefont {Sandvik}\ \emph {et~al.}(2010)\citenamefont {Sandvik}, \citenamefont {Avella},\ and\ \citenamefont {Mancini}}]{Sandvik_2010}%
  \BibitemOpen
  \bibfield  {author} {\bibinfo {author} {\bibfnamefont {A.~W.}\ \bibnamefont {Sandvik}}, \bibinfo {author} {\bibfnamefont {A.}~\bibnamefont {Avella}}, \ and\ \bibinfo {author} {\bibfnamefont {F.}~\bibnamefont {Mancini}},\ }in\ \href {\doibase 10.1063/1.3518900} {\emph {\bibinfo {booktitle} {{AIP} Conference Proceedings}}}\ (\bibinfo  {publisher} {{AIP}},\ \bibinfo {year} {2010})\BibitemShut {NoStop}%
\bibitem [{\citenamefont {Nelson}\ and\ \citenamefont {Kosterlitz}(1977)}]{Nelson_and_Kosterlitz_1977}%
  \BibitemOpen
  \bibfield  {author} {\bibinfo {author} {\bibfnamefont {D.~R.}\ \bibnamefont {Nelson}}\ and\ \bibinfo {author} {\bibfnamefont {J.~M.}\ \bibnamefont {Kosterlitz}},\ }\href@noop {} {\bibfield  {journal} {\bibinfo  {journal} {Phys. Rev. Lett.}\ }\textbf {\bibinfo {volume} {39}},\ \bibinfo {pages} {1201} (\bibinfo {year} {1977})}\BibitemShut {NoStop}%
\bibitem [{\citenamefont {Binder}(1981)}]{Binder1981}%
  \BibitemOpen
  \bibfield  {author} {\bibinfo {author} {\bibfnamefont {K.}~\bibnamefont {Binder}},\ }\href {\doibase 10.1007/BF01293604} {\bibfield  {journal} {\bibinfo  {journal} {Zeitschrift f{\"u}r Physik B Condensed Matter}\ }\textbf {\bibinfo {volume} {43}},\ \bibinfo {pages} {119} (\bibinfo {year} {1981})}\BibitemShut {NoStop}%
\bibitem [{\citenamefont {Binder}(1997)}]{Binder1997}%
  \BibitemOpen
  \bibfield  {author} {\bibinfo {author} {\bibfnamefont {K.}~\bibnamefont {Binder}},\ }\href@noop {} {\bibfield  {journal} {\bibinfo  {journal} {Reports on Progress in Physics}\ }\textbf {\bibinfo {volume} {60}},\ \bibinfo {pages} {487} (\bibinfo {year} {1997})}\BibitemShut {NoStop}%
\bibitem [{\citenamefont {Xu}\ and\ \citenamefont {Ma}(2007)}]{Xu2007}%
  \BibitemOpen
  \bibfield  {author} {\bibinfo {author} {\bibfnamefont {J.}~\bibnamefont {Xu}}\ and\ \bibinfo {author} {\bibfnamefont {H.-R.}\ \bibnamefont {Ma}},\ }\href {\doibase 10.1103/PhysRevE.75.041115} {\bibfield  {journal} {\bibinfo  {journal} {Phys. Rev. E}\ }\textbf {\bibinfo {volume} {75}},\ \bibinfo {pages} {041115} (\bibinfo {year} {2007})}\BibitemShut {NoStop}%
\bibitem [{\citenamefont {Weber}\ and\ \citenamefont {Minnhagen}(1988)}]{Weber1988}%
  \BibitemOpen
  \bibfield  {author} {\bibinfo {author} {\bibfnamefont {H.}~\bibnamefont {Weber}}\ and\ \bibinfo {author} {\bibfnamefont {P.}~\bibnamefont {Minnhagen}},\ }\href {\doibase 10.1103/PhysRevB.37.5986} {\bibfield  {journal} {\bibinfo  {journal} {Phys. Rev. B}\ }\textbf {\bibinfo {volume} {37}},\ \bibinfo {pages} {5986} (\bibinfo {year} {1988})}\BibitemShut {NoStop}%
\bibitem [{\citenamefont {Jiang}(2024)}]{Jiang2024}%
  \BibitemOpen
  \bibfield  {author} {\bibinfo {author} {\bibfnamefont {F.-J.}\ \bibnamefont {Jiang}},\ }\href {\doibase 10.1093/ptep/ptae147} {\bibfield  {journal} {\bibinfo  {journal} {Progress of Theoretical and Experimental Physics}\ }\textbf {\bibinfo {volume} {2024}},\ \bibinfo {pages} {103A02} (\bibinfo {year} {2024})},\ \Eprint {http://arxiv.org/abs/https://academic.oup.com/ptep/article-pdf/2024/10/103A02/60221145/ptae147.pdf} {https://academic.oup.com/ptep/article-pdf/2024/10/103A02/60221145/ptae147.pdf} \BibitemShut {NoStop}%
\bibitem [{\citenamefont {Gupta}\ and\ \citenamefont {Baillie}(1992)}]{gupta1992}%
  \BibitemOpen
  \bibfield  {author} {\bibinfo {author} {\bibfnamefont {R.}~\bibnamefont {Gupta}}\ and\ \bibinfo {author} {\bibfnamefont {C.~F.}\ \bibnamefont {Baillie}},\ }\href@noop {} {\bibfield  {journal} {\bibinfo  {journal} {Phys. Rev. B}\ }\textbf {\bibinfo {volume} {45}},\ \bibinfo {pages} {2883} (\bibinfo {year} {1992})}\BibitemShut {NoStop}%
\bibitem [{\citenamefont {Wojtkiewicz}\ \emph {et~al.}(2023)\citenamefont {Wojtkiewicz}, \citenamefont {Wohlfeld},\ and\ \citenamefont {Ole\ifmmode~\acute{s}\else \'{s}\fi{}}}]{Wojtkiewicz2023}%
  \BibitemOpen
  \bibfield  {author} {\bibinfo {author} {\bibfnamefont {J.}~\bibnamefont {Wojtkiewicz}}, \bibinfo {author} {\bibfnamefont {K.}~\bibnamefont {Wohlfeld}}, \ and\ \bibinfo {author} {\bibfnamefont {A.~M.}\ \bibnamefont {Ole\ifmmode~\acute{s}\else \'{s}\fi{}}},\ }\href@noop {} {\bibfield  {journal} {\bibinfo  {journal} {Phys. Rev. B}\ }\textbf {\bibinfo {volume} {107}},\ \bibinfo {pages} {064409} (\bibinfo {year} {2023})}\BibitemShut {NoStop}%
\end{thebibliography}%

\end{document}